\documentclass{aip-cp}

\usepackage[numbers]{natbib}
\usepackage{rotating}
\usepackage{graphicx}
\usepackage{siunitx}
\DeclareSIUnit\electron{e^{-}}
\usepackage{textgreek}

\usepackage{tikz}
\newcommand\copyrighttext{%
  \footnotesize This article may be downloaded for personal use only. Any other use requires prior permission of the author and AIP Publishing. The following article appeared in AIP Conf. Proc. \textbf{2054}, 060037 (2019) and may be found at https://doi.org/10.1063/1.5084668 .}
\newcommand\copyrightnotice{%
\begin{tikzpicture}[remember picture,overlay]
\node[anchor=south] at (current page.south) {\fbox{\parbox{\dimexpr\textwidth-\fboxsep-\fboxrule\relax}{\copyrighttext}}};
\end{tikzpicture}%
}

\makeatletter
\renewcommand\@biblabel[1]{#1.}
\makeatother

\begin{document}

\title{Hammerhead, an  Ultrahigh Resolution ePix Camera for Wavelength-Dispersive Spectrometers}

\author[aff1]{G.~Blaj\corref{cor1}}
\author[aff1]{D.~Bhogadi}
\author[aff1]{C.~Chang}
\author[aff1]{D.~Doering}
\author[aff1]{C.~Kenney}
\author[aff1]{T.~Kroll}
\author[aff1]{M.~Kwiatkovski}
\author[aff1]{J.~Segal}
\author[aff1]{D.~Sokaras}
\author[aff1]{G.~Haller}


\affil[aff1]{SLAC National Accelerator Laboratory, 2575 Sand Hill Road, Menlo Park, CA 94025, U.S.A.}
\corresp[cor1]{Corresponding author: blaj@slac.stanford.edu}

\maketitle
\copyrightnotice

\begin{abstract}
Wavelength-dispersive spectrometers (WDS) are often used in synchrotron and FEL applications where high energy resolution (in the order of eV) is important. Increasing WDS energy resolution requires increasing spatial resolution of the detectors in the dispersion direction. The common approaches are not ideal: using strip detectors loses the 2D position sensitivity (important for, e.g., background estimation) and the high counting rate; using pixel detectors with a small pitch results in complex charge sharing behaviour and typically have a small size. Developing pixel detectors with high aspect ratio and a small pitch in the wavelength dispersive direction would be ideal, however, it would require a substantial ASIC development. We present a new approach, with a novel sensor design using rectangular pixels with a high aspect ratio (between strips and pixels, further called ``strixels''), and strixel redistribution to match the square pixel arrays of typical ASICs. This results in a sensor area of \SI{17.4x77}{\milli\metre}, with a fine pitch of \SI{25}{\micro\metre} in the horizontal direction resulting in \num{3072} columns and \num{176} rows. The sensors use ePix100 readout ASICs, leveraging their low noise (\SI{43}{\electron}, or \SI{180}{\electronvolt} rms). We present results obtained with a Hammerhead ePix100 camera, showing that the small pitch (\SI{25}{\micro\metre}) in the dispersion direction maximizes performance for both high and low photon occupancies, resulting in optimal WDS energy resolution. The low noise level at high photon occupancy allows precise photon counting, while at low occupancy, both the energy and the subpixel position can be reconstructed for every photon, allowing an ultrahigh resolution (in the order of \SI{1}{\micro\metre}) in the dispersion direction and rejection of scattered beam and harmonics. Using strixel sensors with redistribution and flip-chip bonding to standard ePix readout ASICs results in ultrahigh position resolution ($\sim$\SI{1}{\micro\metre}) and low noise in WDS applications, leveraging the advantages of hybrid pixel detectors (high production yield, good availability, relatively inexpensive) while minimizing development complexity through sharing the ASIC, hardware, software and DAQ development with other versions of ePix cameras.

\end{abstract}

\section{INTRODUCTION}

Wavelength-dispersive spectrometers (WDS) are often used in synchrotron and FEL applications where high energy resolution (in the order of eV) is important \cite{sokaras2012high, sokaras2013seven}. Increasing WDS energy resolution requires increasing spatial resolution of the detectors in the dispersion direction. One approach to increase the resolution is to use small pixel sizes (e.g., \SI{25x25}{\micro\metre} \cite{dinapoli2014monch}), however, the small ASIC size and complex charge sharing are challenging \cite{cartier2014micron}.

Another approach to increase the resolution while maintaining the advantages of pixel detectors over strip detectors (high counting rates, 2D position sensitivity) is to design rectangular pixels with a high aspect ratio (between strips and pixels, further called ``strixels''). Developing new readout ASICs with rectangular pixels is complex. However, existing readout ASICs for hybrid pixel detectors can be easily matched with novel sensors with rectangular strixels where the strixel contacts are redistributed to match the square pixel arrays of typical ASICs.

The Hammerhead camera uses two novel strixel sensors, developed at SLAC, with a total of \num{176x3072} strixels of \SI{100x25}{\micro\metre} redistributed to match regular \SI{50x50}{\micro\metre} ASIC pixel arrays. Each sensor is flip-chip bonded to two ePix100 readout ASICs \cite{markovic2014design, carini2015the}, leveraging their high resolution (\SI{50x50}{\micro\metre}) and low noise performance (\SI{180}{\electronvolt} rms or \SI{43}{\electron} equivalent noise charge, ENC).

We present results obtained with a Hammerhead ePix100 camera, showing that the small pitch (\SI{25}{\micro\metre}) in the dispersion direction maximizes both the subpixel resolution and the energy resolution of charge summing, while yielding good results at high photon occupancies. The low noise level at high photon occupancy allows precise photon counting, while at low occupancy, both the energy and the subpixel position can be reconstructed for every photon, allowing an ultrahigh resolution (in the order of \SI{1}{\micro\metre}) in the dispersion direction and rejection of scattered beam and harmonics.

Using strixel sensors with redistribution and flip-chip bonding to standard ePix readout ASICs results in ultrahigh position resolution ($\sim$\SI{1}{\micro\metre}) and low noise in WDS applications, leveraging the advantages of hybrid pixel detectors (high production yield, good availability, relatively inexpensive) while minimizing development complexity through sharing the ASIC, hardware, software and DAQ development with other versions of ePix cameras \cite{dragone2013epix}.

\section{MATERIALS AND METHODS}

\subsection{High Resolution Sensor}
\begin{figure}[h]
  \centerline {%
    \includegraphics[width =6.5in]{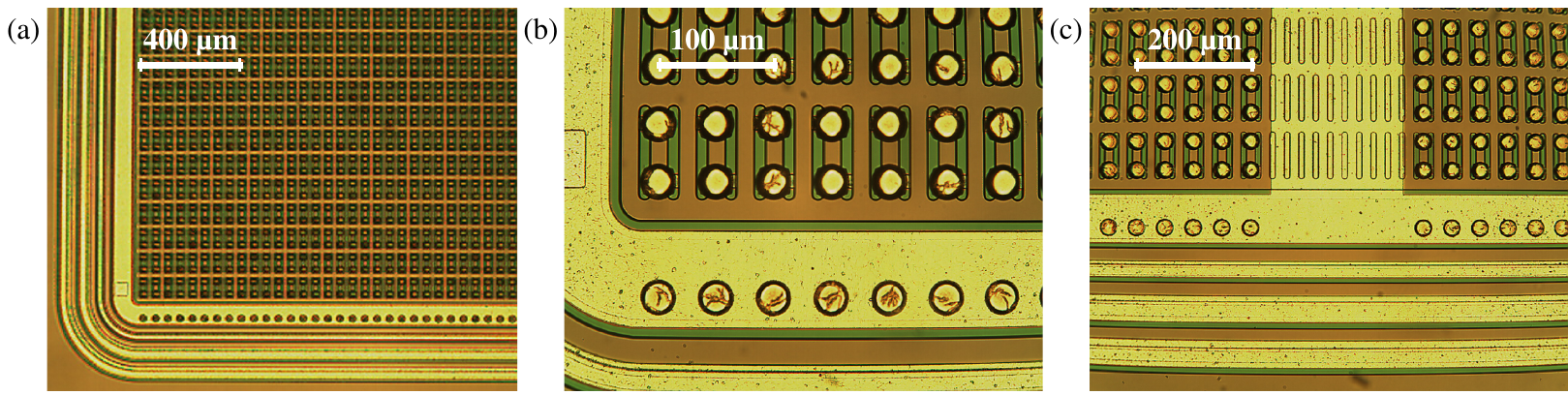}
  }
  \caption{Microscope images of the strixel sensor, showing bump-bond pads (yellow circles, matching read-out ASICs) and implant masks (green vertical lines, matching strixel charge collection area): (a) at first sight, the sensor looks like regular hybrid pixel sensors, with equally spaced bump-bond pads; (b) zooming in reveals the bump-bond pads on a regular \SI{50x50}{\micro\metre} grid; underneath are the implant mask lines (corresponding to \SI{25x100}{\micro\metre} strixels). One strixel row is connected to two rows of bump-bond pads, with the top row connected to the strixels to their right and the bottom row to the left (connections just visible in the image); (c) sensor edge area between two ASICs, with a gap of \SI{250}{\micro\metre}; guard ring structure visible on the bottom.}
  \label{fig1}
\end{figure}

We developed a novel spectroscopic sensor with high aspect ratio pixels (further called ``strixels''), redistributing the contacts of the \SI{100x25}{\micro\metre} strixels to a regular grid of \SI{50x50}{\micro\metre} of flip-chip bonding pads and matching the existing ePix100 readout ASICs \cite{markovic2014design}; details show in Fig.~\ref{fig1}.

Each sensor has a size of \SI{17.6x39.2}{\milli\metre}, with a strixel size of \SI{100x25}{\micro\metre}, resulting in \num{176} rows and \num{1536} columns with a high position resolution along the horizontal axis. Each sensor is flip-chip bonded to two ePix100 ASICs \cite{markovic2014design}, which have a low noise (\SI{43}{\electron} equivalent noise charge, or \SI{180}{\electronvolt} rms) and high signal to noise ratio \cite{blaj2016xray}.

The strixel sensors presented here use typical n-type silicon with a thickness of \SI{500}{\micro\metre} and a resistivity of $\rho=$\SI{10}{\kilo\ohm\centi\metre}; other versions are easy to develop. The development of the novel sensor was far less labor intensive than an alternative development of read-out sensors and ASICs with a matching rectangular shape.

\subsection{Hammerhead ePix100 Camera}
\begin{figure}[h]
  \centerline {%
    \includegraphics[width =6.5in]{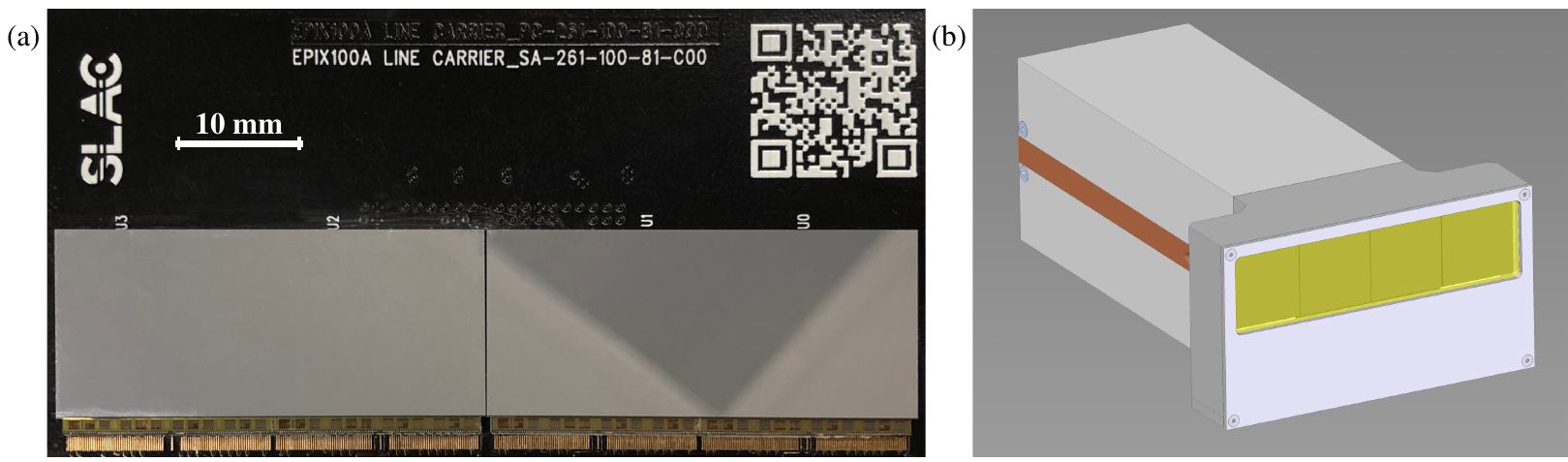}
  }
  \caption{(a) An actual Hammerhead ePix100 module with \num{176x3072} strixels, reusing the ePix100 readout ASICs and ePix camera framework; (b) standard ePix camera design with a modified head to accomodate the wide sensor area.}
  \label{fig2}
\end{figure}

Two sensors are assembled side-by-side in a typical Hammerhead ePix100 module, as shown in Fig.~\ref{fig2}~(a). This results in a module with a sensor area of \SI{17.6x77}{\milli\metre} and \num{3072} columns with a width of \SI{25}{\micro\metre}.

The Hammerhead ePix100 cameras are built around ePix100 hybrid pixel detector modules using the ePix camera platform \cite{dragone2013epix,nishimura2015design}, sharing development of the electronics, mechanical, thermal, vacuum and communication with all other cameras in the family.

\subsection{Experimental Setup and Data Processing}

We used an $^{55}$Fe radioactive source, yielding mainly Mn~K\textalpha{ }at \SI{5.9}{\kilo\electronvolt} and a small amount of Mn~K\textbeta{ }at \SI{6.5}{\kilo\electronvolt}. The source was placed \SI{4}{\centi\metre} from the sensors to illuminate the entire area of the two sensors. For each data set, \num{16384} frames were acquired. The raw data was subsequently processed (dark subtraction, common mode correction, gain correction, droplet reconstruction, subpixel resolution, histogrammed).

All processing was performed with Tensorflow algorithms, taking advantage of recent advances in machine learning (software, GPU support) and yielding 1-2 orders of magnitude faster processing with much simplified code (for details see \cite{blaj2018ultrafast}). This approach enables real time processing of the raw data stream, reducing it to sparse photons and their subpixel resolution (at low photon occupancy), or photon counting data (at high photon occupancy), which can be orders of magnitude more compact than the raw data of integrating detectors.

Note that a gain correction is only useful when extracting the energy dispersive spectrum or subpixel resolution; for photon counting at low occupancy, an average gain is sufficient (as the pixel gain variation is limited to a few percent and has an insignificant effect on photon counting).

\section{RESULTS}

\subsection{High Resolution Imaging and WDS Resolution}
The fine horizontal pitch of \SI{25}{\micro\metre} of the Hammerhead ePix100 camera results in a high resolution of \SI{25}{\micro\metre} (obviously) in the horizontal, i.e., energy dispersive direction, for signals with high photon occupancy. An example of imaging with high horizontal resolution is shown in Fig.~\ref{fig1}(a); the displayed height-to-width ratio corresponds to the physical sensor size; note however the fine pitch in the horizontal direction, corresponding to pixels of \SI{100x25}{\micro\metre} (height x width). The corresponding high resolution spectrum is depicted in Fig.~\ref{fig1}(b).

\begin{figure}[h]
  \centerline {%
    \includegraphics[width =6.5in]{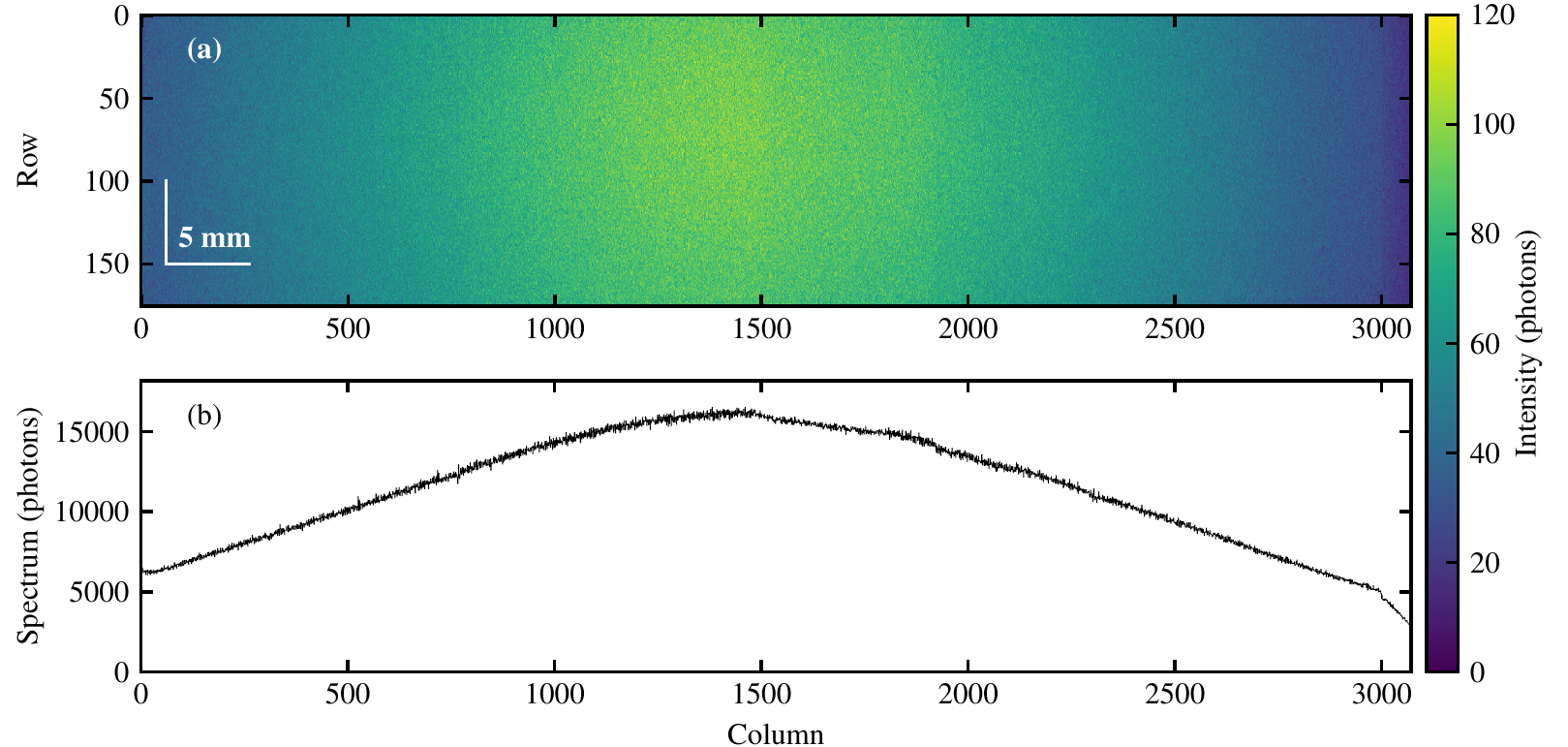}
  }
  \caption{Example of high resolution spectrum acquisition showing a smoothly varying distribution of photons ($^{55}$Fe source a few centimeters from sensor plane); (a) shows a photon counting accumulated image over \num{50000} frames with low photon occupancy; (b) shows the corresponding x projection obtained by summing pixel values in each column, with a representative resolution for any photon occupancy; note the high resolution in the horizontal direction.}
  \label{fig3}
\end{figure}

\subsection{Energy Dispersive Resolution}
The relatively small pixel size (\SI{25}{\micro\metre}) in the horizontal direction results in a relatively large amount of charge sharing between neighboring pixels in the horizontal direction. In the vertical direction, charge sharing is minimized by the large vertical pixel size (of \SI{100}{\micro\metre}). Because charge sharing is mostly due to the horizontal charge sharing, both the energy dispersive spectrum and the subpixel resolution can be calculated with high accuracy.

\begin{figure}[h]
  \centerline {%
    \includegraphics[width =6.5in]{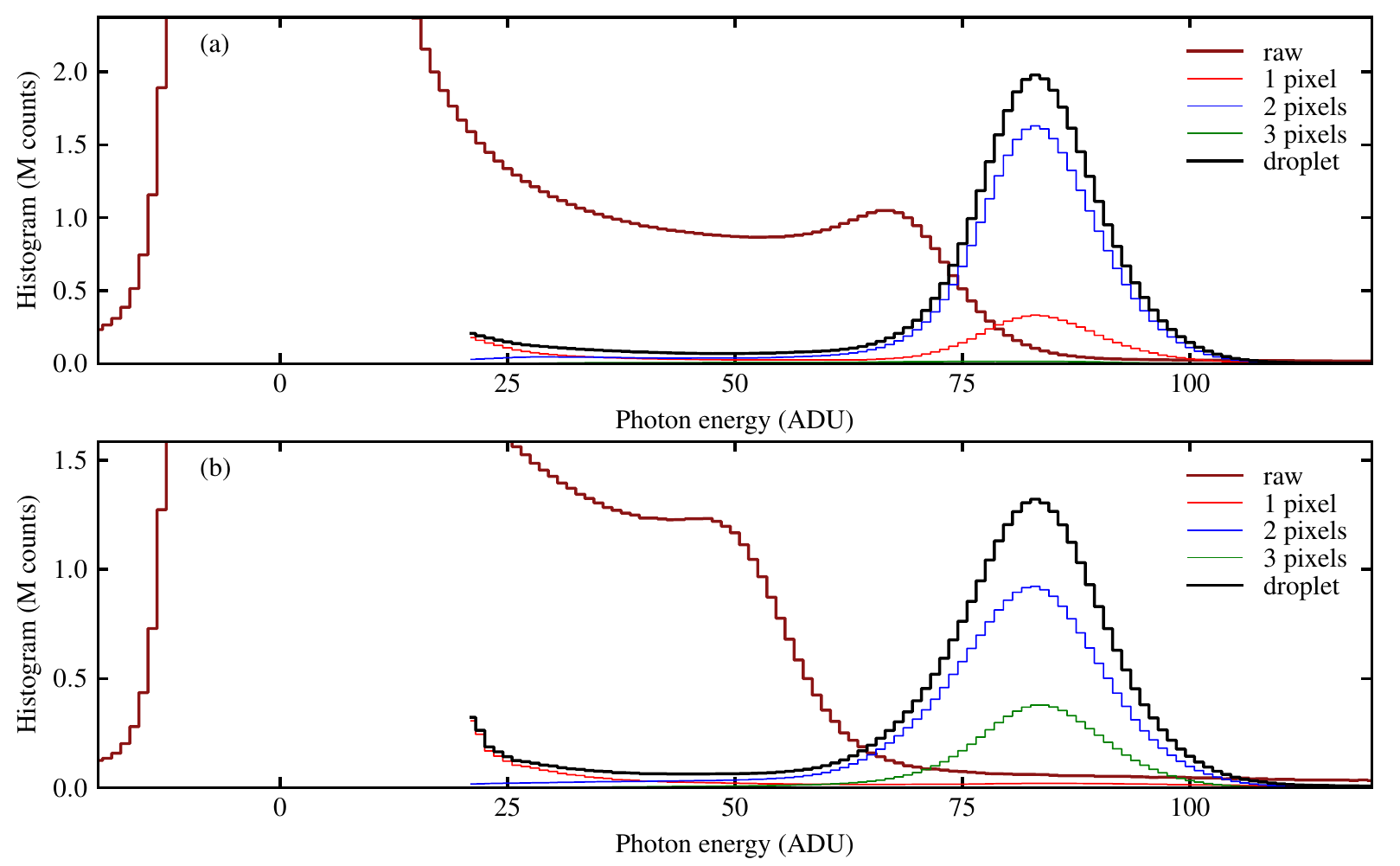}
  }
  \caption{The energy dispersive spectroscopic performance (a) at typical \SI{200}{\volt} bias, and (b) at minimal \SI{100}{\volt} bias. The raw spectrum (thick red line) shows significant charge sharing. The thick black line depicts the reconstructed spectrum, using charge summing and matching the expected \num{83}~ADU gain for single photons. Note the successful reconstruction even with intentionally challenging biasing at \SI{100}{\volt} sensor bias (just over full depletion). Thin lines indicate the contributions from photon charge clouds with a width of \num{1}, \num{2} and \num{3} pixels.}
  \label{fig4}
\end{figure}

The energy dispersive spectrum results are shown in Fig.~\ref{fig4}; the thick red line indicates the raw spectrum and the thick black line shows the result of charge summing (which rectifies the effect of charge sharing). Panel (a) shows the results at a typical bias voltage of \SI{200}{\volt}, while panel (b) shows the results with intentionally maximized charge sharing. Note that in both cases, charge summing rectifies the position of the photon peak at \num{83}~ADU.

The thin lines indicate the contributions of individual photons with charge cloud sizes of \num{1}, \num{2} and \num{3} pixels. At \SI{200}{\volt}, \SI{81}{\percent} of photons are detected in \num{2} pixel clusters and \SI{17}{\percent} in \num{1} pixel clusters.  At \SI{100}{\volt}, \SI{71}{\percent} of photons are detected in \num{2} pixel clusters and \SI{26}{\percent} in \num{3} pixel clusters.

\subsection{Subpixel resolution at low photon occupancy}

\begin{figure}[h]
  \centerline {%
    \includegraphics[width =6.5in]{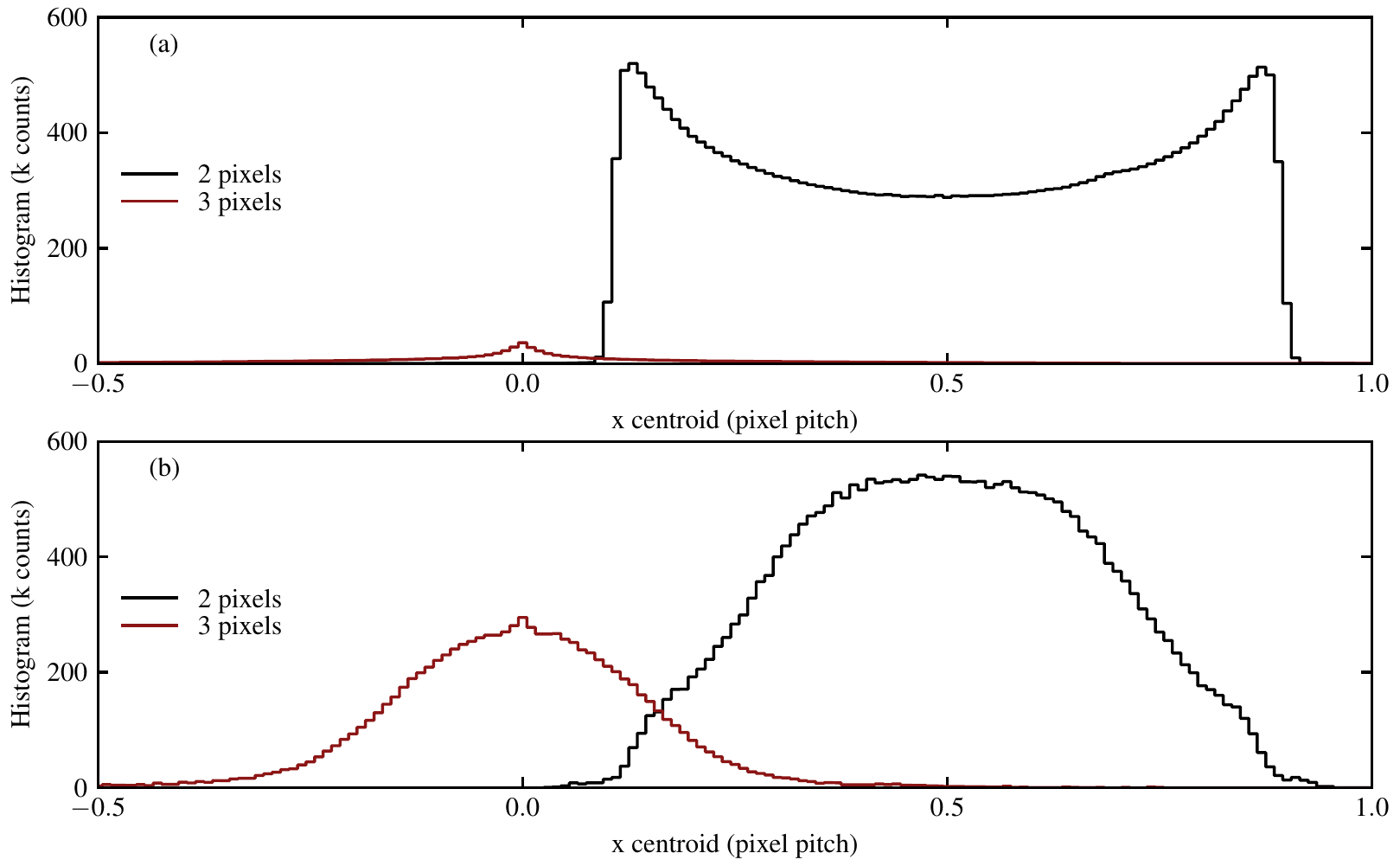}
  }
  \caption{Histogram of subpixel centroids in the x direction with (a) typical \SI{200}{\volt} sensor bias and (b) \SI{100}{\volt} sensor bias; for improved subpixel resolution, a bias of \SI{100}{\volt} is superior (the optimal bias is probably between \SI{100}{\volt} and \SI{200}{\volt}).}
  \label{fig5}
\end{figure}

This results at a typical bias voltage of \SI{200}{\volt} in a high subpixel accuracy for horizontal detection further than \SI{5}{\micro\metre} from the center of the nearest pixel, and no subpixel accuracy within \SI{2.5}{\micro\metre} of the pixel center; in practice, the subpixel position resolution is limited to \SI{5}{\micro\metre} in this case, see Fig.~\ref{fig5}(a).

The average charge cloud size depends on the photon energy, sensor characteristics (thickness, resistivity), and bias voltage (for a discussion, see next subsection); decreasing the bias voltage to \SI{100}{\volt} results in larger photon clouds, Fig.~\ref{fig4}(b), resulting in turn in a high position resolution in the order of \SI{1}{\micro\metre} everywhere, including photons detected near the center of the pixels.
    
\begin{figure}[h]
  \centerline {%
    \includegraphics[width =6.5in]{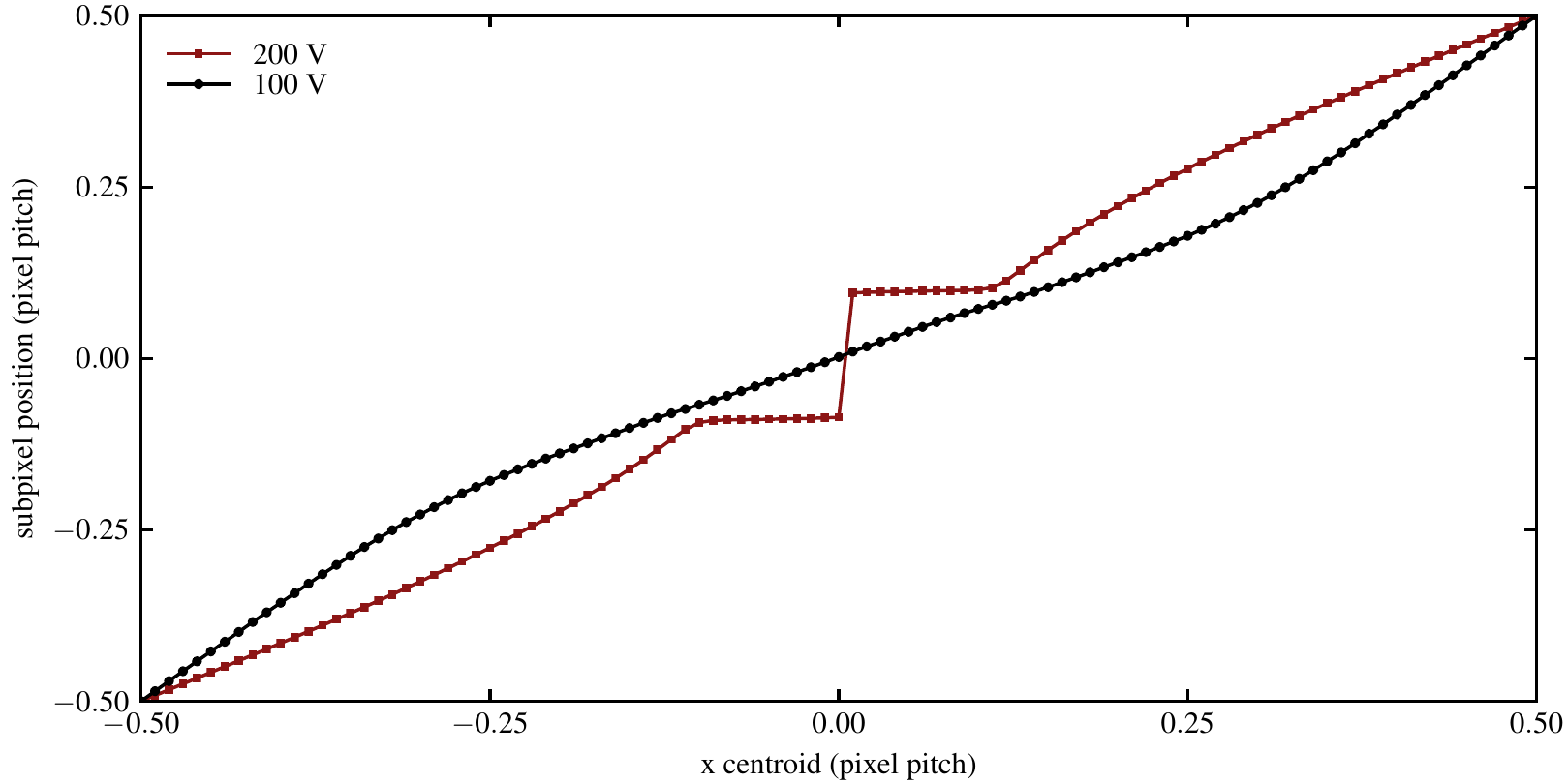}
  }
  \caption{Relation between x centroid and subpixel position; for \SI{200}{V} bias (red line with squares), subpixel resolution is limited to \SI{5}{\micro\metre} within $\pm$\SI{10}{\percent} of pixel center and \SI{1}{\micro\metre} elsewhere; a \SI{100}{\volt} bias (black line with dots) allows \SI{1}{\micro\metre} resolution everywhere.}
  \label{fig6}
\end{figure}

Successful measurements with ultrahigh subpixel resolution (in the order of \SI{1}{\micro\metre}) require optimization of the bias voltage for the photon energy, and a calibration of the relation between centroid positions and corresponding photon positions (using either measurement statistics, see Fig.~\ref{fig6}, or theoretical considerations, see next subsection).

\subsection{Size of Photon Charge Clouds}
Sensors in hybrid pixel detectors are typically reverse-biased p-n junctions; after photon detection and conversion to electron-hole charge clouds, the charge drift can be described by the partial differential equations describing charge drift, diffusion and recombination \cite{blaj2017analytical}. We derive from first principles the factors that influence the size of the charge clouds, which in turn are important for optimizing detector parameters for subpixel resolution.

For a fully depleted sensor, the bias voltage $V$ must be higher than minimum voltage for full sensor depletion $V_D$:
\begin{equation} \label{eq1}
    V \geq V_D=\frac{q_e N d^2}{2 \varepsilon}
\end{equation}
which for the sensor described in this paper is $\sim$\SI{85}{\volt} (calculations and measurements in good agreement). Using the convention in \cite{blaj2017analytical}, with coordinate $x$ denoting the depth of the sensor measured from the origin of real or virtual transition from fully depleted to undepleted region, the front and back planes of the fully depleted sensors are at
\begin{eqnarray}  \label{eq2}
    x_1=\frac{\varepsilon \mu_n \rho_n V}{d} - \frac{d}{2}, x_2=x_1+d.
\end{eqnarray}
The charge drift time from photon detection depth $\xi \in (x_1,x_2)$ to position $x_2$ is
\begin{equation}  \label{eq3}
    t_d(\xi)=\frac{x_2-\xi}{v_s}+\frac{\ln \big(\frac{x_2}{\xi}\big)}{b}
\end{equation}
leading to a lateral charge cloud size due to thermal diffusion
\begin{equation}  \label{eq4}
    \sigma(\xi)=\sqrt{2 D t_d(\xi)}.
\end{equation}
which is a function of sensor bias $V$ and actual photon detection depth in the sensor $\xi$. The distribution of photon detection depths follows the typical exponential decay function
\begin{equation}  \label{eq5}
    I(\xi) = I_0 e^{-\mu(\xi-x_1)}=I_0 e^{-\mu(\xi-x_2+d)}
\end{equation}
where $\mu$ is the linear attenuation coefficient in the sensor (Si) corresponding to the photon energy $E$. Using this distribution allows calculating an average cloud size as a quadratic weighted average of the cloud sizes for all possible detection depths, weighted by their probabilities:
\begin{equation}  \label{eq6}
    \bar{\sigma}^2=\frac{\int_{x_2-d}^{x_2} \sigma^2(\xi) p(\xi) d\xi}{\int_{x_2-d}^{x_2} p(\xi) d\xi} = \frac{2 D \mu}{1-e^{-\mu d}} \int_{x_2-d}^{x_2}\bigg[\frac{x_2-\xi}{v_s}+\frac{\ln\big(\frac{x_2}{\xi}\big)}{b}\bigg] e^{-\mu(\xi-x_2+d)} d\xi
\end{equation}
\begin{equation}  \label{eq7}
    \bar{\sigma}^2=-\frac{1}{\mu v_s}\frac{e^{-\mu d}+\mu d -1}{e^{-\mu d}-1}+\frac{1}{b}\frac{e^{\mu x_2}[Ei(\mu x_1)-Ei(\mu x_2)]+e^{\mu d}\ln(\frac{x_2}{x_1})}{e^{\mu d}-1}
\end{equation}
where $Ei$ is the exponential integral:
\begin{equation}  \label{eq8}
    Ei(x)=-\int_{-x}^{\infty}\frac{e^{-t}}{t} dt
\end{equation}
The average charge cloud size depends thus on the bias voltage $V$ and photon energy $E$ as shown in Eq.~\ref{eq7}.

\section{CONCLUSIONS}

Strixel sensors with redistribution leverage the advances in 2D hybrid pixel detector development, reusing existing pixel readout ASICs with minimal effort and enabling high horizontal resolutions for wave dispersive spectrometer applications. Strixel sensors maintain all advantages of 2D pixel detectors (high counting rate, 2D position sensitive) and  enable a high position accuracy of \SI{25}{\micro\metre} in all conditions.

By using subpixel resolution at low photon occupancies, an ultrahigh horizontal position resolution of \SI{5}{\micro\metre} with default sensor bias of \SI{100}{\volt}, and \SI{1}{\micro\metre} with bias optimization. The high subpixel resolution is enabled by the high signal to noise ratio operation of the ePix100 cameras.

The Hammerhead ePix100 cameras leverage the advantages of hybrid pixel detectors (high production yield, good availability, relatively inexpensive) while minimizing development complexity through sharing the ASIC, hardware, software and DAQ development with other versions of ePix cameras. 

Using strixel sensors with high aspect ratio pixels results in a fine pitch in the energy dispersive direction. By redistributing the flip-chip bonding pads to match the square grid of existing ASICs we obtain a low noise and ultrahigh position resolution (\SI{25}{\micro\metre} at high photon occupancy and \SI{1}{\micro\metre} at low photon occupancy) in WDS applications, leveraging the advantages of hybrid pixel detectors (high production yield, good availability, relatively inexpensive) while minimizing development effort through sharing the ASIC, hardware, software and DAQ development with other versions of ePix cameras \cite{blaj2015future}. Versions for other readout ASICs with different pixel pitches are easy to develop.


\section{ACKNOWLEDGMENTS}
Use of the Stanford Synchrotron Radiation Lightsource, SLAC National Accelerator Laboratory, is supported by the U.S. Department of Energy, Office of Science, Office of Basic Energy Sciences under Contract No. \mbox{DE-AC02-76SF00515}. Publication number SLAC-PUB-17276.



%

\end{document}